\documentclass[oneside,english,12pt,a4]{article}
\usepackage[latin1]{inputenc}
\usepackage{amssymb}
\usepackage{amsmath}
\usepackage{graphicx}
\usepackage{babel}
\usepackage{a4wide}
\begin{document}
\title{Static and dynamic properties of the XXZ chain with long-range interactions}
\author{L. L. Gonçalves$^{a}$\thanks{Email address \textit{lindberg@fisica.ufc.br} (L. L.
Gonçalves).}, L. P. S. Coutinho$^{a}$, J. P. de Lima$^{b}$}
\date{}
\maketitle
\begin{center}
 $^{(a)}$Departamento de Física, Universidade Federal
do Ceará,

Campus do Pici, caixa postal 6030, 60451-970 Fortaleza, Ceará,
Brazil

$^{(b)}$Departamento de Física, Universidade Federal do Piauí,

Campus Ministro Petrônio Portela, 64049-550, Teresina, Piauí,
Brazil
\end{center}
\begin{abstract}
The one-dimensional XXZ model (s=1/2) in a transverse field, with
uniform long-range interactions among the transverse components of
the spins, is studied. The model is exactly solved by introducing
the Jordan-Wigner transformation and the integral Gaussian
transformation. The complete critical behaviour and the critical
surface for the quantum and classical transitions, in the space
generated by the transverse field and the interaction parameters,
are presented. The crossover lines for the various
classical/quantum regimes are also determined exactly. It is shown
that, besides the tricritical point associated with the classical
transition, there are also two quantum critical points: a
bicritical point where the classical second-order critical line
meets the quantum critical line, and a first-order transition
point at zero field. It is also shown that the phase diagram for
the first-order classical/quantum transitions presents the same
structure as for the second-order classical/quantum transitions.
The critical classical and quantum exponents are determined, and
the internal energy, the specific heat and the isothermal
susceptibility,$\chi_{T}^{zz},$ are presented for the different
critical regimes. The two-spin static and dynamic correlation
functions, $<S_{_{j}}^{z}S_{l}^{z}>$, are also presented, and the
dynamic susceptibility, $\chi_{q}^{zz}$($\omega),$is obtained in
closed form. Explicit results are presented at $T=0,$ and it is
shown that the isothermal susceptibility, $\chi_{T}^{zz},$ is
different from the static one, $\chi_{q}^{zz}$($0).$ Finally, it
is shown that, at $T=0,$ the internal energy close to the
first-order quantum transition satisfies the scaling form recently
proposed by Continentino and Ferreira.


\noindent crossover; dynamic properties; quantum phase transition.


\end{abstract}

\bigskip\bigskip\bigskip
\textit{Preprint submitted to PHYSICA A.}

\newpage
\section{{\large Introduction}}
Quantum transitions are induced by quantum fluctuations which occur
in the limit of very low temperatures ($T\rightarrow0$), where they
dominate over thermal fluctuations responsible for inducing classical
transitions {[}1{]}. The Mott transition, between extended and localized
electronic states, is perhaps the best known and studied example of
quantum transitions (metal-insulator transition) {[}2{]}.

The observation of anomalous behaviour in magnetic materials at very
low temperatures has stimulated the study of the quantum transitions
in these systems {[}3,4{]}. In particular, since the transitions occur
at $T=0$, the study of magnetic chains is particularly welcome in
view of the possibility of obtaining exact solutions which allow a
rigorous description. This has been the main motivation for considering
in this work the XXZ chain, with a uniform long-range interaction,
where quantum and classical transitions are allowed and an exact solution
can be obtained. Although the model has already been considered by
Suzuki {[}5{]}, his study was restricted to the analysis of the classical
second-order transition. As it will be shown, besides first-order
classical transitions, the model also presents quantum transitions
of first and second-order. This model is amenable to rigorous study
of the classical/ quantum crossover. Since we are interested in the
complete description of the critical behaviour (quantum and classical
transitions of first and second-order), it will be considered again.

In section 2 we present the solution of the model and obtain the
equation of state. The classical critical behaviour is obtained in
section 3, and the quantum critical one in section 4. The dynamic
correlations in the field direction are studied in section 5 and
in section 6 we present the dynamic susceptibility. Finally, in
section 7, we obtain the critical surface for the quantum and
classical transitions, and the crossover lines separating the
various critical regimes.
\section{{\large Basic results and the equation of state}}
We consider the one-dimensional XXZ model ($s=1/2$, $N$ sites)
with uniform long-range interactions among the $z$ components of
the spins. The Hamiltonian is explicitly given by
\begin{equation}
H=-J\sum_{j=1}^{N}\left(S_{j}^{x}S_{j+1}^{x}+S_{j}^{y}S_{j+1}^{y}\right)-\frac{I}{N}\sum_{j,k=1}^{N}S_{j}^{z}S_{k}^{z}-h\sum_{j=1}^{N}S_{j}^{z},\end{equation}
\noindent where $N$ is the number of sites of the lattice and we
have assumed periodic boundary conditions. By applying the
Jordan-Wigner fermionization {[}6,7{]},
\begin{equation}
S_{j}^{+}=\left[\exp\left(i\pi\sum_{l=1}^{j-1}c_{l}^{\dagger}c_{l}\right)\right]c_{j}^{\dagger};\quad\quad\quad
S_{j}^{-}=c_{j}\left[\exp\left(-i\pi\sum_{l=1}^{j-1}c_{l}^{\dagger}c_{l}\right)\right],\end{equation}
this Hamiltonian can be written in the form
\begin{center}
\begin{equation}
\begin{split}
H&=-\frac{J}{2}\sum_{j=1}^{N}\left(c_{j}^{\dagger}c_{j+1}+c_{j+1}^{\dagger}c_{j}\right)-\left(h-I\right)
\sum_{j=1}^{N}c_{j}^{\dagger }c_{j}- \\
&-\frac{I}{N}\sum_{j,l=1}^{N}c_{j}^{\dagger }c_{j}c_{l}^{\dagger }c_{l}+%
\frac{N}{2}\left( h -\frac{I}{2}\right) .
\end{split}
\end{equation}
\end{center}
The partition function is then given by
\begin{equation}
\begin{split}
Z_{N}&=\exp\left[-\frac{\beta
N}{2}\left(h-\frac{I}{2}\right)\right]Tr\left\{
\exp\left[\frac{\beta
J}{2}\sum_{j=1}^{N}\left(c_{j}^{\dagger}c_{j+1}+c_{j+1}^{\dagger}c_{j}\right)\right]\right.\times
\\
&\left.\times\exp\left[\beta\left(h-I\right)\sum_{j=1}^{N}c_{j}^{\dagger}c_{j}+\frac{\beta
I}{N}\sum_{j,l=1}^{N}c_{j}^{\dagger}c_{j}c_{l}^{\dagger}c_{l}\right]\right\},
\end{split}
\end{equation}
where $\beta$ =$1/(k_{B}T)$ and $T$ is the temperature. It should
be noted that the term with long-range interactions commutes with
the Hamiltonian. This decomposition allows the introduction, in
the second exponential, of the Gaussian transformation {[}8{]}
\begin{equation}
\exp\left(a^{2}\right)=\frac{1}{\sqrt{2\pi}}\int_{-\infty}^{\infty}\exp\left(-\frac{x^{2}}{2}+\sqrt{2}ax\right)dx,\end{equation}
such that the partition function can be written in the integral
representation
\begin{equation}
\begin{split}
Z_{N}&=\exp\left[-\frac{N}{2}\left(\overline{h}-\frac{1}{2}\right)\right]\sqrt{\frac{N}{2\pi}}\int_{-\infty}^{\infty}\exp\left(-\frac{N\overline{x}^{2}}{2}\right)\times
\\
& \times Tr\left\{
\exp\left[\frac{\overline{J}}{2}\sum_{j=1}^{N}\left(c_{j}^{\dagger}c_{j+1}+c_{j+1}^{\dagger}c_{j}\right)+\left(\overline{h}-I+\sqrt{2\overline{I}\overline{x}}\right)\sum_{j=1}^{N}c_{j}^{\dagger}c_{j}\right]\right\}
d\overline{x},
\end{split}
\end{equation}
where $\overline{x}\equiv
x/\sqrt{N},$ $\overline{J}\equiv\beta J,$ $\overline{h}\equiv\beta
h\,\,$and$\,\,$$\overline{I}\equiv\beta I.$

Introducing the canonical transformation
\begin{equation}
c_{k}^{\dagger}=\frac{1}{\sqrt{N}}\sum_{j=1}^{N}\exp\left(-ikj\right)c_{j}^{\dagger},\end{equation}
with $k=2\pi n/N$, $n=1,2,3,..N,$ eq. (6) can finally be written
in the form
\begin{equation}
Z_{N}=C(\beta)\int_{-\infty}^{\infty}\exp\left(-\frac{N^{2}}{2}\right)\zeta(\overline{x})d\overline{x},\end{equation}
where
\begin{equation}
C(\beta)\equiv\sqrt{\frac{N}{2\pi}}\exp\left[-\frac{N}{2}\left(\overline{h}-\frac{1}{2}\right)\right],\quad\zeta(\overline{x})\equiv
Tr\left\{
\exp\left[\sum_{k}\overline{\epsilon}_{k}(\overline{x})c_{k}^{\dagger}c_{k}\right]\right\}
,\end{equation} and
\begin{equation}
\overline{\epsilon}_{k}(\overline{x})=\beta\epsilon_{k}(\overline{x})\equiv\beta\left(J\cos(k)+h-I+\sqrt{\frac{2I}{\beta}}\overline{x}\right).\end{equation}

In the thermodynamic limit, $N\longrightarrow\infty,$ the
partition function can be calculated by Laplace's method {[}9{]},
\begin{equation}
Z_{N}=\frac{\exp\left[-\frac{N}{2}\left(\overline{h}-\frac{1}{2}\right)+Ng(\overline{x}_{0})\right]}{\left|g^{\prime\prime}(\overline{x}_{0})\right|^{\frac{1}{2}}},\quad\text{with}\quad
g^{\prime}(\overline{x}_{0})=0,g^{\prime\prime}(\overline{x}_{0})<0,\end{equation}
and where
\begin{equation}
g(\overline{x})=-\frac{\overline{x}_{0}^{2}}{2}+\frac{1}{N}\sum_{k}\ln\left[1+\exp\left(\overline{\epsilon}_{k}\right)\right].\end{equation}

The Helmholtz free energy is given by
\begin{equation}
\mathfrak{F}_{N}=\frac{N}{2}\left(h-\frac{I}{2}\right)-Nk_{B}Tg(\overline{x}_{0})+\frac{k_{B}T}{2}\ln\left|g^{\prime\prime}(\overline{x}_{0})\right|,\end{equation}
where $\overline{x}_{0}$ is expressed in terms of the
magnetization per site $M^{^{z}},$
\begin{center}
\begin{equation}
\begin{split}
M^{z}=\frac{1}{N}\sum_{j}<S_{j}^{z}>&=\frac{1}{N}\sum_{k}<c_{k}^{\dagger}c_{k}>-\frac{1}{2} \\
&=\frac{1}{N}\sum_{k}\frac{1}{1+\exp\left[-\overline{\epsilon}_{k}(\overline{x}_{0})\right]}-\frac{1}{2}.
\end{split}
\end{equation}
\end{center}
Using eqs. (11) and (12), in the thermodynamic limit, we have
\begin{equation}
M^{z}+\frac{1}{2}=\frac{\overline{x}_{0}}{\sqrt{2\overline{I}}}.\end{equation}
From this result it follows that the functional of the Helmholtz
free energy per site is finally given by
\begin{equation}
\mathfrak{f}=\frac{h}{2}-\frac{k_{B}T}{\pi}\int_{0}^{\pi}\ln\left\{
1+\exp\left[\overline{\epsilon}_{k}(M^{z})\right]\right\}
dk+IM^{z}\left(M^{z}+1\right),\end{equation}
 where
\begin{equation}
\overline{\epsilon}_{k}(M^{z})=-\overline{J}\cos(k)-\overline{h}-2\overline{I}M^{z}.\end{equation}

The equation of state is obtained from eq. (16) by imposing the
conditions
\begin{equation}
\frac{\partial\mathfrak{f}}{\partial
M^{z}}=0,\quad\text{with}\quad\frac{\partial^{2}\mathfrak{f}}{\partial
M^{z2}}>0.\end{equation} We then have
\begin{equation}
M^{z}=\frac{1}{2\pi}\int_{0}^{\pi}\tanh\left[\frac{\cos(k)+\Gamma+2rM^{z}}{2\widetilde{T}}\right]dk,\end{equation}
where $\,\, r\equiv I/J,$ $\Gamma\equiv h/J$ and
$\widetilde{T}\equiv k_{B}T/J.$
\section{{\large The classical critical behaviour }}
For $I<0$, the system is totally frustrated. Consequently, it does
not present any classical critical behaviour. On the other hand, for
$I>0$, there is a classical critical behaviour; the system undergoes
first and second-order transitions at a finite temperature.

The classical phase diagram is obtained from the equation of
state, by considering $h=0.$ The second-order transitions are
determined by taking in this equation the limit
$M^{^{z}}\longrightarrow0$, and the first-order classical
transitions are obtained by solving the equations
\begin{center}
\begin{subequations}
\begin{align}
M_{t}^{z}-\frac{1}{2\pi}\int_{0}^{\pi}\tanh\left[\frac{\cos(k)+2rM_{t}^{z}}{2\widetilde{T_{t}}}\right]dk
& =  0,\\
\mathfrak{f}(M_{t}^{z})&=\mathfrak{f}(0),
\end{align}
\end{subequations}
\end{center}
where $M_{t}^{z}$ $\,\,$ is the finite value of the magnetization
at the transition. In passing we would like to note that eq.(19)
reproduces the known results in the limits $J=0$ {[}10{]} and
$I=0$ {[}11{]}.
 The first-order critical line meets the
second-order line at the tricritical point, which can be obtained
by imposing the condition that the second and fourth derivatives
of the functional of the Helmholtz energy go to zero as
$M^{^{z}}\longrightarrow0$. This leads to the set of equations
\begin{center}\begin{subequations}\begin{align}
\frac{r_{tr}}{2\pi\widetilde{T}_{tr}}\int_{0}^{\pi}sech^{2}\left[\frac{\cos(k)}{2\widetilde{T}_{tr}}\right]dk-1&=0,\\
\int_{0}^{\pi}\left\{
-2\tanh^{2}\left[\frac{\cos(k)}{2\widetilde{T}_{tr}}\right]sech^{2}\left[\frac{\cos(k)}{2\widetilde{T}_{tr}}\right]+sech^{4}\left[\frac{\cos(k)}{2\widetilde{T}_{tr}}\right]\right\}
dk&=0\mathbf{,}
\end{align}
\end{subequations}
\end{center}
from which we have $\widetilde{T}_{tr}=0,37716...$and
$r_{tr}=1,39815...$
 There is a critical value $r_{c}$ , which is a
lower bound for the first-order line, below which the system does
not present classical behaviour; it corresponds to a first-order
quantum transition. This point is determined by considering
$M_{t}^{z}=1/2$ and the limit $\widetilde{T}\rightarrow0,$ in
eqs.(20a) and (20b), which gives $r_{c}=4/\pi.$

In a finite field, the critical line associated with the
second-order transitions is obtained by requiring that the minimum
of the functional of the Helmholtz free energy is triply
degenerate. From eq.(16), this leads to the set of equations
\begin{center}\begin{subequations}\begin{align}
\frac{1}{2\pi}\int_{0}^{\pi}\tanh\left[\frac{\overline{\epsilon}_{k}(M_{cr}^{z})}{2}\right]dk-M_{cr}^{z}&=0,\\
\frac{r}{2\pi\widetilde{T}}\int_{0}^{\pi}sech^{2}\left[\frac{\overline{\epsilon}_{k}(M_{cr}^{z})}{2}\right]dk-1&=0,\\
\frac{r^{2}}{\widetilde{T}^{2}}\int_{0}^{\pi}\tanh\left[\frac{\overline{\epsilon}_{k}(M_{cr}^{z})}{2}\right]sech^{2}\left[\frac{\overline{\epsilon}_{k}(M_{cr}^{z})}{2}\right]dk&=0,
\end{align}
\end{subequations}
\end{center}
where $r_{cr}$ is restricted to the interval $0<r_{cr}<r_{tr}$.

 The classical phase diagram shown in Fig.1 can
be constructed by solving eqs.(20-22). As it can be seen, below
the tricritical point there is a curve of pseudo second-order
transitions which correspond to metastable solutions. This
classical phase diagram is very similar to analogous phase diagram
of an Ising model with short and long-range interactions {[}12{]},
as it will be shown below, is characterized by the same
(mean-field) exponents.

The critical exponent $\beta$ associated with the magnetization
can be determined from eqs. (19), (21) and (22). It is found to be
equal to $1/2$ for the second-order transitions and along the
classical critical line, at nonzero field, and equal to $1/4$ at
the tricritical point$.$

The classical transitions can also be characterized by the
non-analytical behaviour of other thermodynamic functions. In
particular, the internal energy and specific heat present
non-analytical behaviour which are related to the order of the
transition {[}10{]}.
\begin{figure}[t]
\begin{center}
\includegraphics[width=0.8\textwidth]{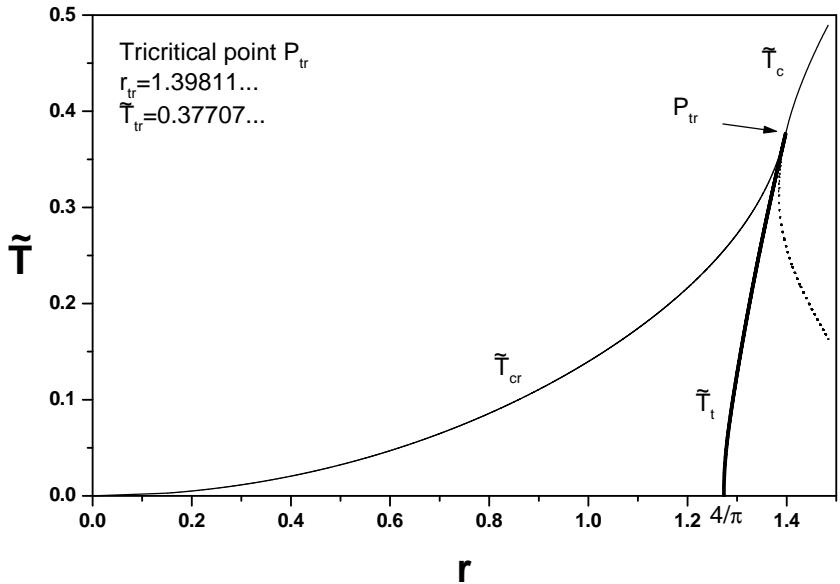}
\end{center}
\caption{Phase diagram for the classical transitions as a function
of the strength of the\ long-range interaction $r$($r=I/J).$
$\widetilde{T}_{t}$( $\widetilde{T}_{t}=k_{B}T_{t}/J)$ identifies
the first-order boundary, $\widetilde{T}_{c}$(
$\widetilde{T}_{c}=k_{B}T_{c}/J)$ the second-order critical line,
P$_{tr}$ the tricritical point and\ the
$\widetilde{T}_{cr}$($\widetilde{T}_{cr}=k_{B}T_{cr}/J)$ the
critical line at nonzero field.} \label{fig1}
\end{figure}
In order to analyze these behaviours let us consider the internal
energy $U\equiv\langle H\rangle.$ Using eqs. (3) and (7), we can
write
\begin{equation}
U=\sum_{k}\varepsilon_{k}\langle
n_{k}\rangle-\frac{I}{N}\sum_{kk^{\prime}}\langle
n_{k}n_{k^{\prime}}\rangle+\frac{N}{2}\left(h-\frac{I}{2}\right),\end{equation}
where $\varepsilon_{k}$ and $\langle n_{k}\rangle$ are given by
\begin{center}
\begin{subequations}\begin{align}
\varepsilon_{k}&=-J\cos(k)-h+I,\\
\langle
n_{k}\rangle&=\frac{1}{1+\exp\left[\overline{\epsilon}_{k}(M^{z})\right]}.
\end{align}
\end{subequations}
\end{center}
By using Wick's theorem {[}13{]}, the average $\langle
n_{k}n_{k^{\prime}}\rangle$ can be written in the form
\begin{equation}
\langle n_{k}n_{k^{\prime}}\rangle=\langle n_{k}\rangle\langle
n_{k^{\prime}}\rangle+\langle
c_{k}^{\dagger}c_{k^{\prime}}\rangle\langle
c_{k}c_{k^{\prime}}^{\dagger}\rangle+\langle
c_{k}^{\dagger}c_{k^{\prime}}^{\dagger}\rangle\langle
c_{k}c_{k^{\prime}}\rangle.\end{equation} \noindent From this
result, and bearing in mind that $\langle
c_{k}^{\dagger}c_{k^{\prime}}^{\dagger}\rangle=\langle
c_{k}c_{k^{\prime}}\rangle=0,$ we can write the internal energy as

\begin{eqnarray}
U & = & \sum_{k}\varepsilon_{k}\langle n_{k}\rangle+\frac{N}{2}\left(h-\frac{I}{2}\right)-\nonumber \\
 &  & -\frac{I}{N}\sum_{kk^{\prime}}\langle n_{k}\rangle\langle n_{k^{\prime}}\rangle-\frac{I}{N}\sum_{k}\langle n_{k}\rangle+\frac{I}{N}\sum_{k}\langle n_{k}\rangle^{2}.\end{eqnarray}
In the thermodynamic limit, we can write the internal energy per
site $u\left(u=U/N\right)$ in the explicit form
\begin{equation}
u=\frac{1}{\pi}\int_{0}^{\pi}\frac{\varepsilon_{k}}{1+\exp\left[\overline{\epsilon}_{k}(M^{z})\right]}-IM^{z}\left(M^{z}+1\right)+\frac{1}{2}\left(h-I\right),\end{equation}
which reproduces the known results for $J=0$ {[}14{]} and $I=0$
{[}15{]}.

The specific heat, given by $c_{h}=(\partial u/\partial
T)_{h}=-(\partial u/\partial\beta)_{h}/k_{B}T^{2}$, can be written
as
\begin{equation}
\begin{split}
\frac{c_{h}}{k_{B}}&=\frac{1}{\pi\widetilde{T}^{2}}\left\{
\int_{0}^{\pi}\frac{\left(\cos(k)+\Gamma\right)^{2}\exp\left(-\overline{\epsilon_{k}}\right)}{\left[1+\exp\left(-\overline{\epsilon_{k}}\right)\right]^{2}}dk+\right.\\
&+\left(4rM^{z}+\frac{2r}{\widetilde{T}}\left.\frac{\partial
M^{z}}{\partial\beta}\right|_{h}\right)\int_{0}^{\pi}\frac{\left(\cos(k)+\Gamma\right)\exp\left(-\overline{\epsilon_{k}}\right)}{\left[1+\exp\left(-\overline{\epsilon_{k}}\right)\right]^{2}}dk+\\
&\left.+2rM^{z}\left(2rM^{z}+\frac{2r}{\widetilde{T}}\left.\frac{\partial
M^{z}}{\partial\beta}\right|_{h}\right)\int_{0}^{\pi}\frac{\exp\left(-\overline{\epsilon_{k}}\right)}{\left[1+\exp\left(-\overline{\epsilon_{k}}\right)\right]^{2}}dk\right\}
,
\end{split}
\end{equation}
where
\begin{equation}
\left.\frac{\partial
M^{z}}{\partial\beta}\right|_{h}=\frac{\frac{1}{\pi}\int\limits
_{0}^{\pi}\frac{\left(\cos(k)+\Gamma+2rM^{z}\right)\exp\left(-\overline{\epsilon_{k}}\right)}{1+\exp\left(-\overline{\epsilon_{k}}\right)}dk}{1-\frac{2r}{\pi\widetilde{T}}\int\limits
_{0}^{\pi}\frac{\exp\left(-\overline{\epsilon_{k}}\right)}{1+\exp\left(-\overline{\epsilon_{k}}\right)}dk},\end{equation}
and, naturally, the values of $M^{z}$ also satisfy the equation of
state.

 Typical results for the internal energy and the
specific heat are shown in Fig. 2. As it can be seen in Fig. 2(a),
where we have a classical first-order phase transition, the
internal energy is discontinuous at the critical temperature
(there is a latent heat associated with the transition) and the
specific heat is singular but finite at this point . As shown in
Fig. 2(b), where we have second-order transitions, the specific
heat is also finite at the transition. However, it can be shown
that it diverges at the tricritical point. The critical exponent
$\alpha$ associated with the specific heat is equal to zero along
the second-order transition line and $1/2$ at the tricritical
point, which, as already pointed out, is the mean-field result.

The critical behaviour of the internal energy and specific heat
along the critical line, is identical to the one at the
tricritical point. Therefore, as expected , the exponent $\alpha$
is equal to $1/2.$

The classical critical behaviour can also be characterized by the
isothermal susceptibility $\chi_{T}^{zz}$. From the equation of
state, eq.(19), we have
\begin{equation}
\chi_{T}^{zz}=\frac{\frac{1}{4\pi\widetilde{T}}\int_{0}^{\pi}sech^{2}\left(\frac{\cos(k)+\Gamma+2rM^{z}}{2\widetilde{T}}\right)dk}{1-\frac{r}{2\pi\widetilde{T}}\int_{0}^{\pi}sech^{2}\left(\frac{\cos(k)+\Gamma+2rM^{z}}{2\widetilde{T}}\right)dk}.\end{equation}
The results for the susceptibility, shown in Fig. 3, are in
agreement with typical mean-field behaviour at first and
second-order classical transitions. For the first-order
transition, shown in Fig. 3(a), the susceptibility is singular but
finite at the transition temperature.
\begin{figure}[h]
\begin{center}
\includegraphics[width=0.8\textwidth]{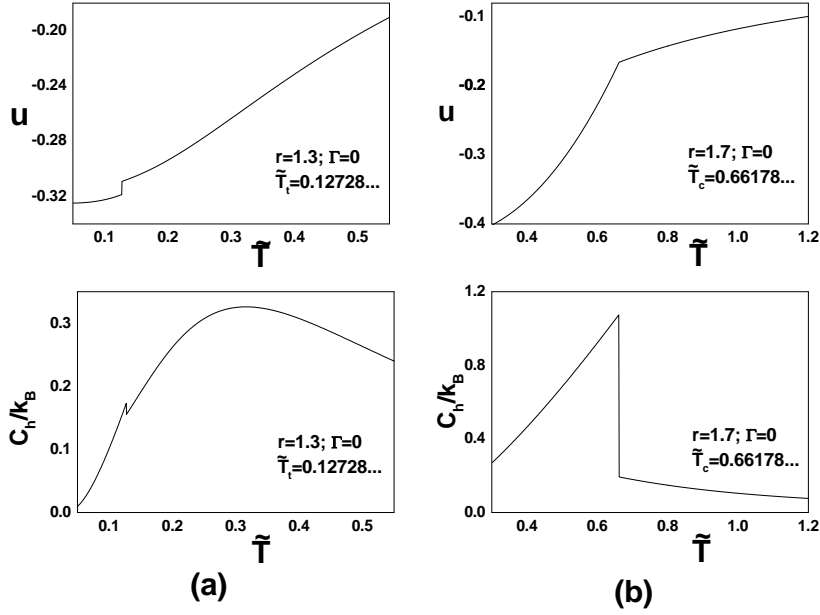}
\end{center}
\caption{Internal energy (u) and specific heat c$_{h}$, at zero
field, as a function of temperature. (a) For $r$=1.3 ($r=I/J)$
where the system undergoes a classical first-order transition. (b)
For\ $r$=1.7 where the system undergoes a classical second-order
transition.} \label{fig2}
\end{figure}
\begin{figure}[h]
\begin{center}
\includegraphics[width=14cm]{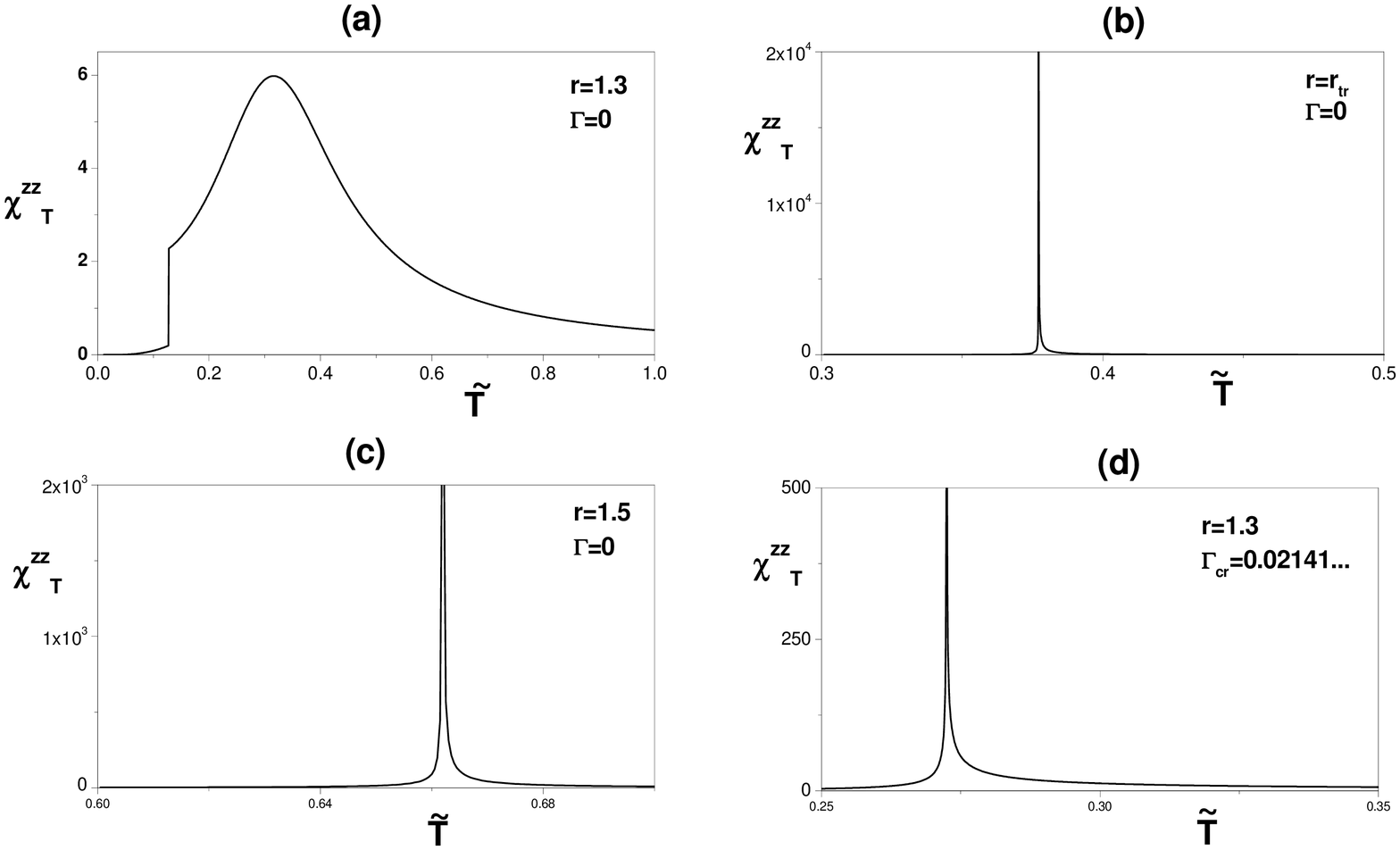}
\end{center}
\caption{Isothermal susceptibility $\chi_{T}^{zz},$ at zero field,
as a function of temperature. (a) \ For $r$=1.3 ($r=I/J)$\ where
the system undergoes a classical first-order transition. (b) At
the tricritical point. (c) For $r$=1.5 where the system undergoes
a classical second-order transition at zero field. (d) At the
critical field $\Gamma_{cr}$\ ($\Gamma_{cr}=h_{cr}/J,$
$\Gamma_{cr}=0.02141...)$ for $r$=1.3.} \label{fig3}
\end{figure}
On the other hand, along the
second-order transition line as well as along the critical line,
and at the tricritical point, it diverges, although with different
critical exponents $\gamma.$ These divergences, which are
signatures of second-order transitions, are shown in Figs. 3(b),
3(c) and 3(d). We have calculated $\gamma$ and have found that it
is equal to 1 at the tricritical point and along the second-order
transition line, whereas it is equal to 1/2 along the nonzero
field critical line. It should be remarked that the exponents
$\alpha,\beta$ and $\gamma$ satisfy the Rushbrook scaling
relation, $\alpha+2\beta+\gamma=2$ {[}10{]}, and that the
classical critical behaviour belongs to the same universality
class of the Ising model with short and long-range
interactions{[}12{]}\textbf{.}
\section{{\large The quantum critical behaviour }}
The quantum phase diagram is determined by from the functional of
the Helmholtz free energy, in the limit $T\rightarrow0.$ For $J>0$
(or $J<0)$ eq. (16) can be written as {[}16{]}
\begin{equation}
f=\frac{h}{2}-\frac{1}{\pi}\int_{0}^{\varphi}\left[J\cos(k)+h+2IM^{z}\right]dk+IM^{z}\left(M^{z}+1\right),\end{equation}
where $\varphi$is given by
\begin{equation}
\varphi=\arccos\left[-\left(\Gamma+2rM^{z}\right)\right].\end{equation}
Also, we have the explicit form
\begin{equation}
f=\frac{h}{2}-\frac{1}{\pi}\left[\sin(\varphi)+\left(h+2IM^{z}\right)\varphi\right]+IM^{z}\left(M^{z}+1\right).\end{equation}
As in the classical region, the equation of state is obtained by
imposing the conditions shown in eq.(18). Then we have
\begin{equation}
M^{z}=\frac{1}{\pi}\arccos\left[-\left(\Gamma+2rM^{z}\right)\right]-\frac{1}{2},\end{equation}
which corresponds to the equation of state, provided the condition
of minimum of$\,\,\,\, f$ is satisfied.

For $I>0$ the system presents first-order transitions induced by
the field, which are determined by the condition

\begin{equation}
f(M_{t}^{z})=f\left(\frac{1}{2}\right).\end{equation} Together
with eq. (34) this leads to the system of equations
\begin{center}
\begin{subequations}
\begin{align}
M_{t}^{z}+\frac{1}{2}-\frac{1}{\pi}\arccos\left[-\left(\Gamma_{t}+2rM_{t}^{z}\right)\right]&=0,\\
-\frac{1}{\pi}\left[\sin(\varphi_{t})+\left(\Gamma_{t}+2rM_{t}^{z}\right)\varphi_{t}\right]+rM_{t}^{z}\left(M_{t}^{z}+1\right)+\Gamma_{t}+\frac{r}{4}&=0,
\end{align}
\end{subequations}
\end{center}
where
\begin{equation}
\varphi_{t}=\arccos\left[-\left(\Gamma_{t}+2rM_{t}^{z}\right)\right].\end{equation}

This first-order quantum line meets the classical first-order line
in the plane $T=0,$ at$\,\,\Gamma=0$ and $r=4/\pi.$ This result
is easily verified by making $M_{t}^{z}=0$ in eq.(36b)$.$ This line
also meets the second-order line at $r=0$ and $\Gamma=1.$ At this
point it also meets the critical classical line, so that it is called
a bicritical point since it corresponds to the intersection of two
second-order critical lines. The phase diagram for these quantum transitions
is shown in Fig.4 {[}16{]}.
\begin{figure}[t]
\begin{center}
\includegraphics[width=0.7\textwidth]{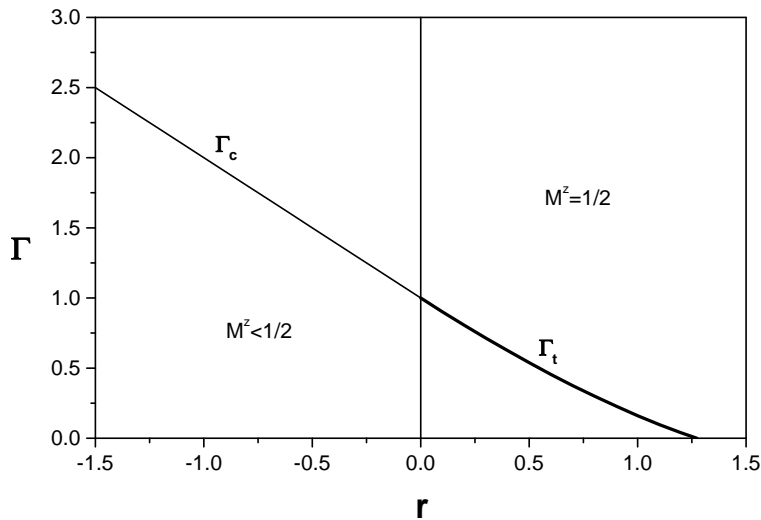}
\end{center}
\caption{Phase diagram for the quantum transitions as a function
of the strength of the\ long-range interaction $r$ ($r=I/J)$.
$\Gamma_{t}$ identifies the first-order transitions and\
$\Gamma_{c}$ the second-order transitions.} \label{fig4}
\end{figure}

The magnetization as a function of the field is shown in Fig.5(a),
where we can see the different types of critical behaviour. For $r<0,$
the magnetization is continuous, since we have second-order transitions,
and it is discontinuous for $r>0,$ where we have first-order transitions.
As in the classical behaviour, there are hysteresis cycles associated
with these quantum first-order transitions, whose limits are obtained
by determining the values of the fields at which the metastable states
disappear.
\begin{figure}[t]
\begin{center}
\includegraphics[width=12cm]{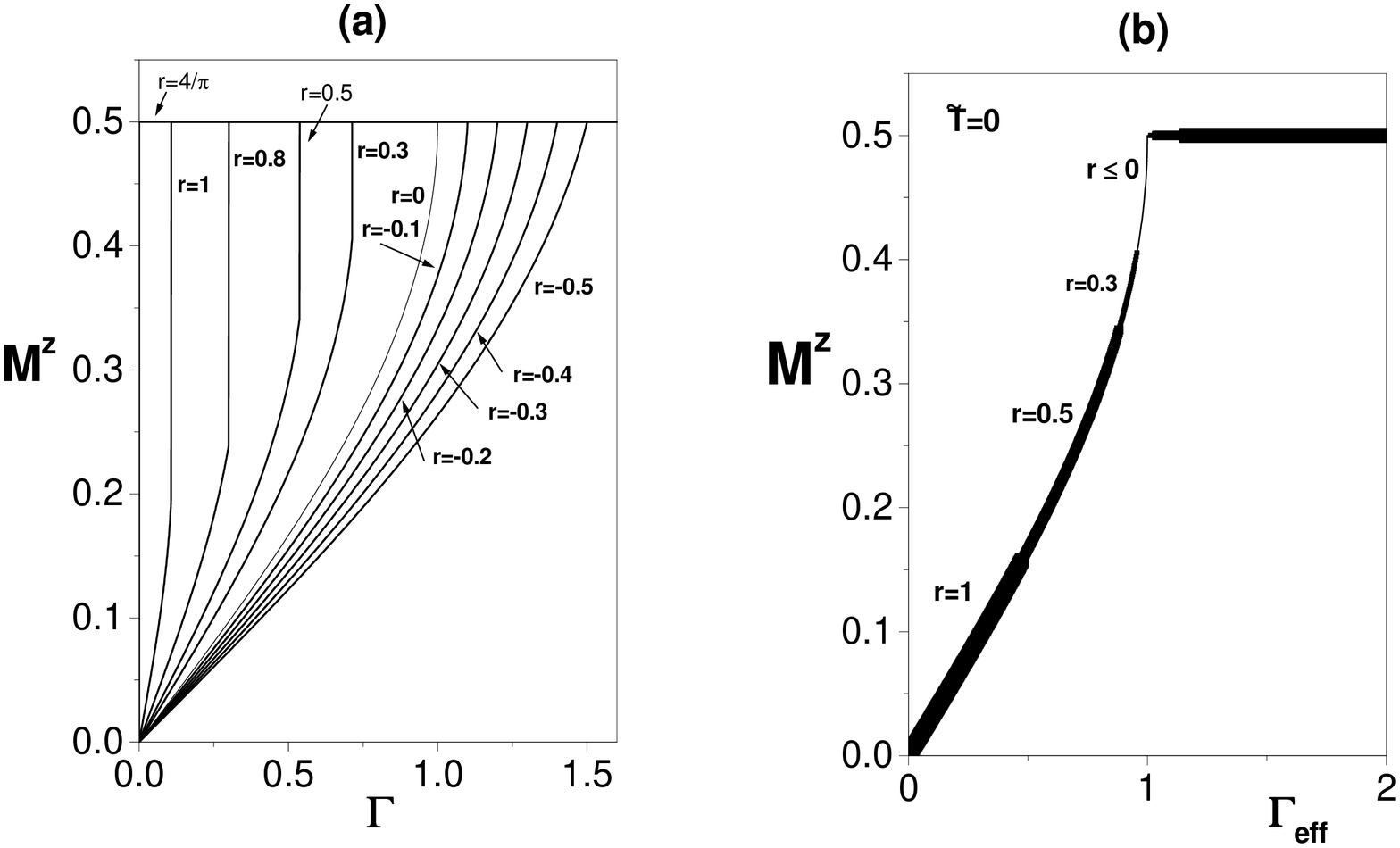}
\end{center}
\caption{(a) Magnetization as a function of
$\Gamma$($\Gamma=h/J$), at $(\widetilde{T}=k_{B}T/J)$
$\widetilde{T}=0$ and for different values of $r$($r=I/J)$, in the
regions where the system undergoes first ($r>0)$ and second-order
($r\leqslant0)$ quantum transitions. (b) Universal curve for the
magnetization as a function of the effective field $\Gamma_{eff}$
$(\Gamma_{eff}=\Gamma+2rM^{z})$ for $\,\, r>0$ and $r\leqslant0$.}
\label{fig5}
\end{figure}
In order to analyze the critical behaviour of the system for
$r<0,$ where it undergoes second-order quantum transitions, we
define the appropriate order parameter {[}17{]}
\begin{equation}
\widetilde{M^{z}}\equiv\frac{1}{2}-M^{z},\end{equation} which goes
to zero ($\widetilde{M^{z}}\rightarrow0^{+}$) as the field
approaches to the critical field. Therefore, by expanding eq. (35)
up to second-order in this parameter, we obtain
\begin{equation}
\frac{\pi^{2}}{2}\left(\widetilde{M^{z}}\right)^{2}-2r\widetilde{M^{z}}=\Gamma_{c}-\Gamma\geq0,\end{equation}
where $\Gamma_{c}=1-r.$ This result reduces to the known one for
$r=0$, namely, $\Gamma_{c}=1.$ The critical exponent $\beta$ is
easily obtained from the previous expression, and it is given by
$\beta=1/2$ for $r=0$ and $\beta=1$ for $r\neq0$ {[}16{]}.
\begin{figure}[h]
\begin{center}
\includegraphics[width=14cm]{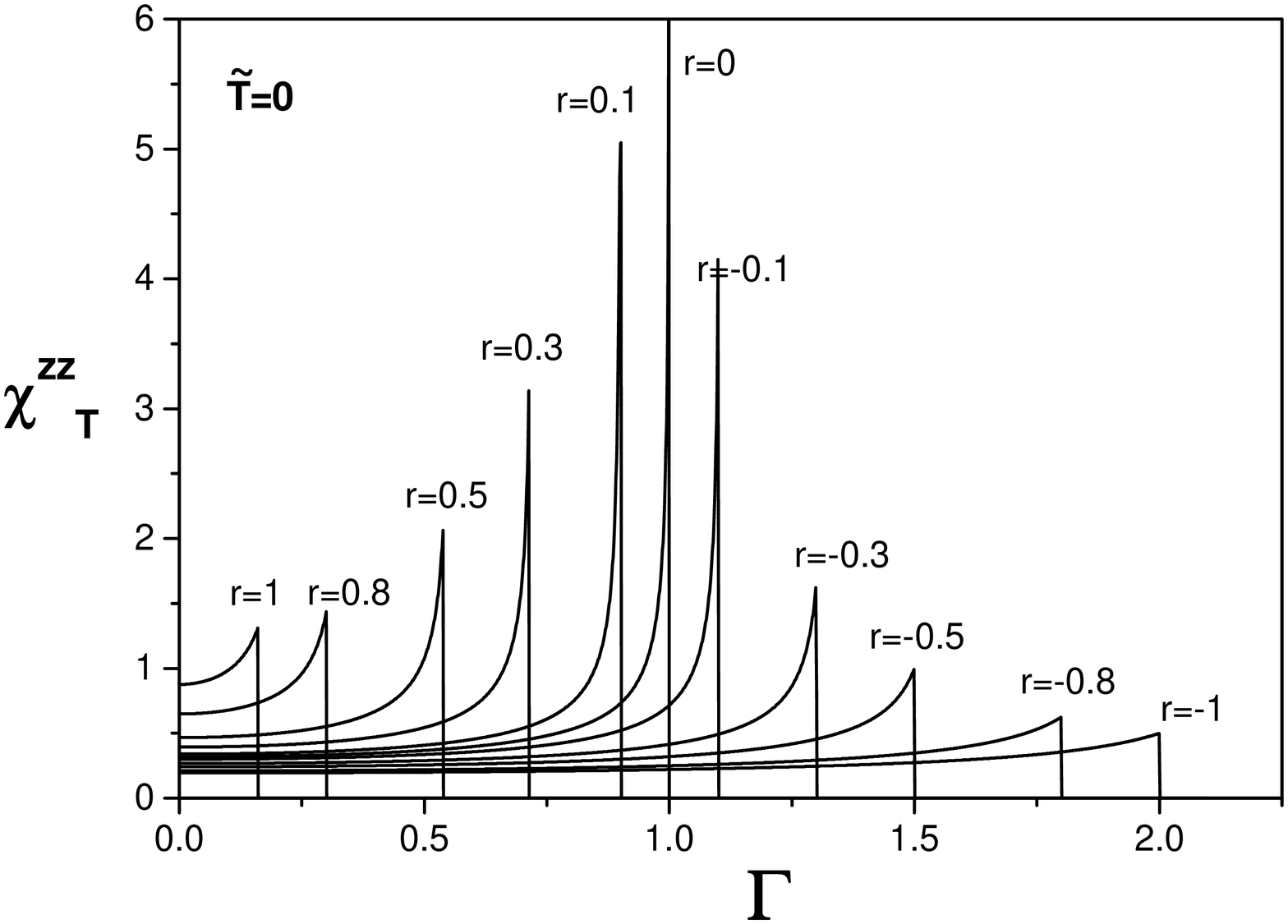}
\end{center}
\caption{Isothermal susceptibility $\chi_{T}^{zz},$ at
$\widetilde{T}=0$ $(\widetilde{T}=k_{B}T/J)$ $,$ as a function of
the field $\Gamma$ ($\Gamma=h/J$) for different values of
$r$($r=I/J)$.} \label{fig6}
\end{figure}
In passing we would like to mention that if we treat the term
$2IM^{z}$ as an external field $h^{\prime}$, as pointed out by
Suzuki {[}5{]}, we can define an effective field
$h_{eff}=h+h^{\prime}.$ In terms of this new field variable the
magnetization becomes a universal curve, even when we have
first-order transitions. The collapse of the magnetization curves,
shown in Fig.5(a), is presented in Fig. 5(b) and could also be
obtained directly from the equation of state. This rather
important result clearly suggests that the quantum critical
behaviour is associated with the effective field
$h_{eff}=h+2M^{z}I^{\prime}$ $(\Gamma_{eff}\equiv
h_{eff}/J=\Gamma+2M^{z}r),$ since along the critical line the
magnetization is saturated.

In contrast to the behaviour in the classical transitions, the
isothermal susceptibility, obtained from eq. (34), and given by
\begin{equation}
\chi_{T}^{zz}=-\frac{\frac{1}{\pi}\frac{1}{\sqrt{1-(\Gamma+2rM^{z})}}}{1-\frac{2r}{\pi}\frac{1}{\sqrt{1-(\Gamma+2rM^{z})}}},\quad\text{or}\quad\Gamma<\Gamma_{c},\end{equation}
diverges at the quantum second-order transition, for $r=0$ only.
This can be seen in Fig. 6, where the isothermal susceptibility,
shown for different values of $r,$ is finite for $r\neq0.$ This
immediately implies that the critical exponent $\gamma,$ defined
as $\chi_{T}^{zz}\sim|\Gamma_{c}-\Gamma|^{\gamma}$, is equal to
zero, for $r\neq0,$ and equal to 1/2, for $r=0.$ Bearing in mind
that, at $T=0,$ $\alpha$ is identical to $\gamma,$ it can be shown
that the exponents $\alpha,\beta$ and $\gamma$ also satisfy the
Rushbrook scaling relation {[}10{]}. It should be noted that the
previous expression can also be obtained by taking
$\widetilde{T}\rightarrow0$in eq. (30).
Finally, it is worth to
mention that, by using eqs. (33), (34) and (36), we can determine
the internal energy $u$ at $T=0.$ Since it is equal to the
Helmholtz free energy, $u=f$ is a continuous function of $h$
independently of the order of the transition. However, at the
first-order transition, its derivative ($du/dh)$ is discontinuous.
Explicitly, close to the critical field $h_{t},$ we can write
\begin{center}
\begin{subequations}
\begin{align}
f(h^{-})&\cong f(h_{t})+M_{t}\left|h-h_{t}\right|,\quad\text{for}\quad h<h_{t},\\
f(h^{+})&\cong
f(h_{t})-\frac{\left|h-h_{t}\right|}{2},\quad\text{for}\quad
h>h_{t},
\end{align}
\end{subequations}
\end{center}
which satisfy the scaling relations recently proposed by
Continentino and Ferreira {[}18{]} for first-order quantum
transitions, with a critical exponent $\alpha$=1.
\section{{\large The dynamic correlation} $\langle
S_{j}^{z}(t)S_{l}^{z}(0)\rangle$} The dynamic correlation $\langle
S_{j}^{z}(t)S_{l}^{z}(0)\rangle$ is defined as
\begin{equation}
{\langle}S_{j}^{z}(t)S_{l}^{z}(0){\rangle=\langle}\exp\left(-iHt\right)S_{j}^{z}\exp\left(iHt\right)S_{l}^{z}{\rangle,}\end{equation}
where $H$ is the Hamiltonian of the system. Since the long-range
interaction term commutes with $H,$ the dynamic correlation can be
written as
\begin{equation} \langle
S_{j}^{z}(t)S_{l}^{z}(0){\rangle=}\langle\exp\left(-iH^{\prime}t\right)S_{j}^{z}\exp\left(iH^{\prime}t\right)S_{l}^{z}\rangle,\end{equation}
where $H^{\prime}$ is given by
\begin{equation}
H^{\prime}=-J\sum_{j=1}^{N}\left(S_{j}^{x}S_{j+1}^{x}+S_{j}^{y}S_{j+1}^{y}\right)-h\sum_{j=1}^{N}S_{j}^{z}S_{j+1}^{z}.\end{equation}
In terms of fermion operators, and introducing the Fourier
transform, we can write $H^{\prime}$ in the form
\begin{equation}
H^{\prime}=\sum_{k}\lambda_{k}c_{k}^{\dagger}c_{k}+\frac{Nh}{2},\end{equation}
where
\begin{equation}
\lambda_{k}=-J\cos(k)-h.\end{equation} From these results, it
follows that
\begin{center}
\begin{subequations}
\begin{align}
c_{k}(t)&=\exp\left(-i\lambda_{k}t\right)c_{k},\\
c_{k}^{\dagger}(t)&=\exp\left(i\lambda_{k}t\right)c_{k}^{\dagger}.
\end{align}
\end{subequations}
\end{center} Therefore, in terms of these operators, we have
\begin{equation}
S_{j}^{z}(t)=\frac{1}{N}\sum_{kk^{\prime}}\exp\left[ij(k-k^{\prime})\right]\exp\left(i\lambda_{k}t\right)\exp\left(-i\lambda_{k^{\prime}}t\right)c_{k}^{\dagger}c_{k^{\prime}}-\frac{1}{2}.\end{equation}
From this expression, the dynamic correlation is written as
\begin{equation}
\begin{split}
\left\langle S_{j}^{z}(t)S_{l}^{z}(0)\right\rangle &=\frac{1}{N^{2}}\sum_{kk^{\prime}qq^{\prime}}\exp\left[ij(k-k^{\prime})\right]\exp\left[i(\lambda_{k}-\lambda_{k^{\prime}})t\right]\times \\
&\times\exp\left[il(q-q^{\prime})\right]\left\langle
c_{k}^{\dagger}c_{k^{\prime}}c_{q}^{\dagger}c_{q^{\prime}}\right\rangle
-\frac{1}{N}\sum_{k^{\prime}}\left\langle
c_{k}^{\dagger}c_{k}\right\rangle +\frac{1}{4}.
\end{split}
\end{equation}

Introducing the Gaussian transformation and using Wick's theorem
{[}13{]}, the previous expression can be written in the final form
\begin{equation}
\begin{split}
\langle S_{j}^{z}(t)S_{l}^{z}(0)\rangle
&=\left[\frac{1}{N}\sum_{k}\left\langle n_{k}\right\rangle
-\frac{1}{2}\right]^{2}+\\
&+\frac{1}{N^{2}}\sum_{kk^{\prime}}\left\{
\left[\exp\left[ik(j-l)\right]\exp\left(i\lambda_{k}t\right)\left\langle
n_{k}\right\rangle \right]\right.\times,\\
&\times\left[\exp\left[-ik^{\prime}(j-l)\right]\exp\left(-i\lambda_{k^{\prime}}t\right)(1-\langle
n_{k^{\prime}}\rangle)\right.\times\\
&\times\left.\left.\exp\left[-ik^{\prime}(j-l)\right]\exp\left(-i\lambda_{k^{\prime}}t\right)(1-\langle
n_{k^{\prime}}\rangle)\right]\right\}
\end{split}
\end{equation}
where$n_{k}$is given by eq. (24b).

The static correlation $\langle S_{j}^{z}(0)S_{l}^{z}(0)\rangle$
is obtained from the previous expression. After some straightforward
algebraic manipulations {[}19{]}, can be written as
\begin{equation}
\begin{split}
\langle S_{j}^{z}S_{l}^{z}\rangle&=\frac{1}{4}\left\{
\left[\frac{1}{N}\sum_{k}\tanh\left(\frac{\overline{\epsilon_{k}}}{2}\right)\right]^{2}-\right.\\
&\left.-\left[\frac{1}{N}\sum_{k}\cos\left[k\left(j-l\right)\right]\tanh\left(\frac{\overline{\epsilon_{k}}}{2}\right)\right]^{2}\right\}
.
\end{split}
\end{equation}

For large separations, the correlation tends to the square of the
magnetization. At $T=0,$ for $r+\Gamma<1$ and $r<0,$ the direct
correlation, $\rho^{z}(j-l)=\langle
S_{j}^{z}S_{l}^{z}\rangle-\langle S_{j}^{z}\rangle^{2},$ can be
expressed as
\begin{equation}
\begin{split}
\rho^{z}(j-l)&=-\left\{
\frac{\sin\left[(j-l)\arccos(\Gamma_{eff})\right]}{\pi(j-l)}\right\}
^{2}\\
&=\frac{\exp\left[2i(j-l)\arccos(\Gamma_{eff})\right]+\exp\left[-2i(j-l)\arccos(\Gamma_{eff})\right]-2}{4\pi^{2}(j-l)^{2}}.
\end{split}
\end{equation}
From this expression, and following Lima and Gon\c{c}alves
{[}17{]}, we can define an analytical extension for the scaling
form of the direct correlation,
\begin{equation}
\rho(n)=\frac{F(in/\xi)}{n^{p}},\end{equation} where
$p=d+z-2+\eta.$ From this\textbf{\ {}}result\textbf{\ {}}and
eq.(52), we obtain
\begin{equation}
\left(\xi\right)^{-1}=2\arccos(\Gamma),\end{equation}
which gives
at the critical point, where $\xi\rightarrow\infty,$
$\Gamma_{eff}^{c}=1$ ($\Gamma_{eff}=\Gamma_{c}+r).$ Using the
scaling relation
$\xi\sim\left|\Gamma_{eff}-\Gamma_{eff}^{c}\right|^{-\nu},$ we
obtain $\nu=1$ for $r\neq0,$ whereas $\nu$=$1/2$ for $r=0.$ The
main implication of this result is that the dynamical exponent
$z,$obtained from the exponent relation $\nu(z+d)=2-\alpha$
{[}20{]}, is $z=$1 for $r\neq0$ and $z=$2 for $r=0.$ Consequently,
the system presents a non-universal critical dynamical behaviour.
Also it should be noted that these spatial oscillations have been
observed in the XXZ model with short-range interactions only
{[}21{]}.

The real and imaginary parts of $\langle S_{j}^{z}(t)S_{l}^{z}(0)\rangle$
can be explicitly written in the form
\begin{equation}
\begin{split}
Re\langle
S_{j}^{z}(t)S_{l}^{z}(0)\rangle&=\left[\frac{1}{N}\sum_{k}\left\langle
n_{k}\right\rangle -\frac{1}{2}\right]^{2}+\left\{
\left[\frac{1}{N}\sum_{k}\cos\left[k(j-l)\right]\cos(\lambda_{k}t)\left\langle
n_{k}\right\rangle \right]\times\right.\\
&\left.\times\left[\frac{1}{N}\sum_{k^{\prime}}\cos\left[k^{\prime}(j-l)\right]\cos(\lambda_{k^{\prime}}t)(1-\langle
n_{k^{\prime}}\rangle)\right]\right\} +\\
&+\left\{
\left[\frac{1}{N}\sum_{k}\cos\left[k(j-l)\right]\sin(\lambda_{k}t)\left\langle
n_{k}\right\rangle \right]\times\right.\\
&\left.\times\left[\frac{1}{N}\sum_{k^{\prime}}\cos\left[k^{\prime}(j-l)\right]\sin(\lambda_{k^{\prime}}t)(1-\langle
n_{k^{\prime}}\rangle)\right]\right\} ,
\end{split}
\end{equation}
and
\begin{equation}
\begin{split}
Im\langle
S_{j}^{z}(t)S_{l}^{z}(0)\rangle&=\left[\frac{1}{N}\sum_{k}\cos\left[k(j-l)\right]\sin(\lambda_{k}t)\left\langle
n_{k}\right\rangle \right]\times\\
&\times\left[\frac{1}{N}\sum_{k^{\prime}}\cos\left[k^{\prime}(j-l)\right]\cos(\lambda_{k^{\prime}}t)(1-\langle
n_{k^{\prime}}\rangle)\right]+\\
&-\left[\frac{1}{N}\sum_{k}\cos\left[k(j-l)\right]\cos(\lambda_{k}t)\left\langle
n_{k}\right\rangle \right]\times\\
&\times\left[\frac{1}{N}\sum_{k^{\prime}}\cos\left[k^{\prime}(j-l)\right]\sin(\lambda_{k^{\prime}}t)(1-\langle
n_{k^{\prime}}\rangle)\right].
\end{split}
\end{equation}
From these expressions we immediately conclude that, at
$T=\infty,$ the dynamic correlation is real and that, for
arbitrary temperatures, $Im\langle
S_{j}^{z}(t)S_{l}^{z}(0)\rangle\rightarrow0$ and $Re\langle
S_{j}^{z}(t)S_{l}^{z}(0)\rangle\rightarrow\langle
S_{j}^{z}\rangle^{2}$ as $t\rightarrow\infty.$ For classical
second-order transitions and zero field, this result is shown in
Fig.7 for the autocorrelation function, which is similar to the
one at the tricritical point, along the critical line and when we
have first-order classical transitions.
\begin{figure}
\begin{center}
\includegraphics[width=0.7\textwidth]{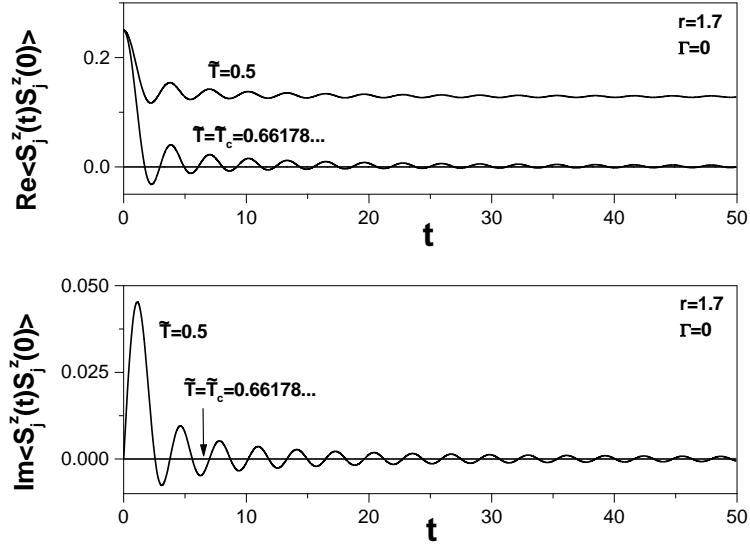}
\end{center}
\caption{Real and \ imaginary parts of the autocorrelation
function $\langle S_{j}^{z}(t)S_{j}^{z}(0)\rangle$ as a function
of $t$, for $\Gamma=0,$ $r=1.7$ ($\Gamma=h/J,$ $r=I/J),$ at the
second-order transition temperature
$\widetilde{T}_{c}=0.66178.(\widetilde{T}=k_{B}T/J)$ and at
$\widetilde{T}=0.5.$} \label{fig7}
\end{figure}
It also should be noted that in the region where the system does
not order the imaginary part goes to zero. The results for
second-order quantum transitions are shown in Fig.8 and present
essentially the same behaviour observed for the classical
transitions irrespective of the order of the transition.
\begin{figure}
\begin{center}
\includegraphics[width=0.7\textwidth]{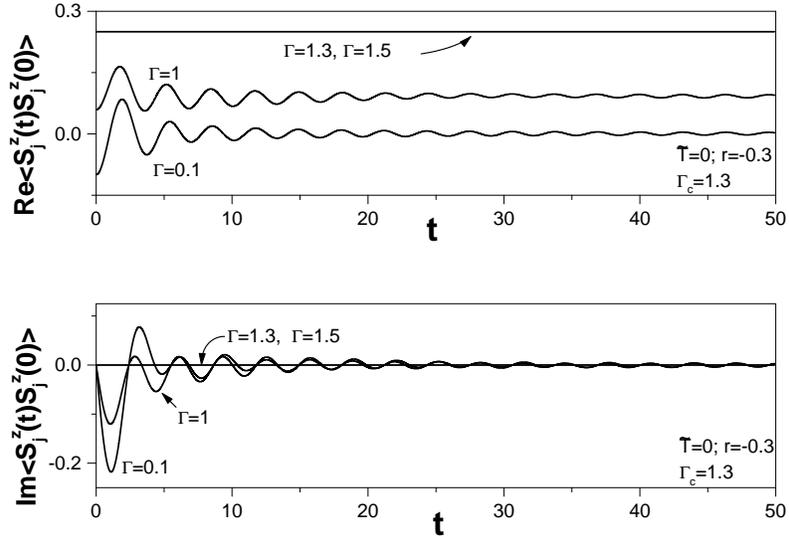}
\end{center}
\caption{Real and \ imaginary parts of the autocorrelation
function $\langle S_{j}^{z}(t)S_{j}^{z}(0)\rangle$ as a function
of $t$, at $\widetilde{T}=0$ $(\widetilde{T}=k_{B}T/J)$ and
$r=-0.3$ ($r=I/J),$ for $\Gamma=1.5;1.0;0.1$ ($\Gamma=h/J$), and
at the second-order quantum transition ($\Gamma_{c}=1.3).$}
\label{fig8}
\end{figure}
\section{{\large The longitudinal dynamic susceptibility }}
The dynamic susceptibility $\chi_{q}^{zz}$($\omega$) is obtained
by considering the time Fourier transform of the two-spin
correlation $\langle S_{j}^{z}(t)S_{l}^{z}(0)\rangle$, which is
given by
\begin{equation}
\langle
S_{j}^{z}S_{l}^{z}\rangle_{\omega}=\frac{1}{2\pi}\int_{-\infty}^{\infty}\langle
S_{j}^{z}(t)S_{l}^{z}(0)\rangle\exp\left(i\omega
t\right)dt.\end{equation} From this equation we obtain
\begin{equation}
\langle S_{j}^{z}S_{l}^{z}\rangle_{\omega}=\delta(\omega)\langle
S_{j}^{z}\rangle^{2}+\frac{1}{N^{2}}\sum_{kk}^{\prime}\exp\left[i(k+k^{\prime})(j-l)\right]\left\langle
n_{k}\right\rangle \left(1-\left\langle n_{
k^{\prime}}\right\rangle
\right)\delta(\omega+\epsilon_{k}-\epsilon_{k^{\prime}}).\end{equation}

Introducing the spatial Fourier transform,
\begin{equation}
\langle
S_{q}^{z}S_{-q}^{z}\rangle_{\omega}=\frac{1}{N}\sum_{j-l}\exp\left[-iq(j-l)\right]\langle
S_{j}^{z}S_{l}^{z}\rangle_{\omega},\end{equation}
we obtain
\begin{equation}
\langle S_{q}^{z}S_{-q}^{z}\rangle_{\omega}=\delta(\omega)\langle
S_{q=0}^{z}\rangle^{2}+\frac{1}{N}\sum_{k}\delta(\omega+\epsilon_{k}-\epsilon_{k-q})\langle
n_{k}\rangle\left(1-\langle n_{k-q}\rangle\right).\end{equation}

From this expression, we can write the susceptibility {[}22{]},
\begin{equation}
\begin{split}
\chi_{q}^{zz}(\omega)&=-2\pi\ll S_{q}^{z};S_{-q}^{z}\gg\\
&=-2\pi\frac{1}{2\pi}\int_{-\infty}^{\infty}\frac{[\exp\left(\beta\omega^{\prime}\right)-1]\exp\left(-\beta\omega^{\prime}\right)\langle
S_{q}^{z}S_{-q}^{z}\rangle_{\omega^{\prime}}}{\omega-\omega^{\prime}}d\omega^{\prime},
\end{split}
\end{equation}
which leads to
\begin{equation}
\chi_{q}^{zz}(\omega)=-\frac{1}{N}\sum_{k}\frac{\langle n_{k}\rangle-\left\langle n_{k-q}\right\rangle }{\omega+\epsilon_{k}-\epsilon_{k-q}}.\end{equation}

The static susceptibility, $\chi_{q}^{zz}(0),$\ is obtained from
the previous expression by considering $\omega=0$ and, in
particular, $\chi_{0}^{zz}(0)$ is determined by taking the limit
$q\rightarrow0,$ which gives, in the thermodynamic limit, the
result
\begin{equation}
\chi_{0}^{zz}(0)=\frac{1}{4\pi\widetilde{T}}\int_{0}^{\pi}sech^{2}\left(\frac{\cos(k)+\Gamma+2rM^{z}}{2\widetilde{T}}\right)dk,\end{equation}
which is different from the isothermal susceptibility
$\chi_{T}^{zz}$ shown in eq. (30).

Differently from the isothermal susceptibility, $\chi_{q}^{zz}(0)$
does not present any infinite singularity at the critical classical
point. At the first-order classical transition, $\chi_{q}^{zz}(0)$
is discontinuous irrespective of the wave-vector, and presents a cuspid
at the second-order classical transition ($\widetilde{T}=\widetilde{T}_{c})$.

At the quantum first-order transition points ($\Gamma=\Gamma_{t}),$
$\chi_{q}^{zz}(0)$ is discontinuous, but finite, for any wave-vector.
However, at the second-order quantum transitions ($\Gamma=\Gamma_{c}),$
$\chi_{q}^{zz}(0)$ diverges at the critical point for $q=0$ only.

The real and imaginary parts of $\chi_{q}^{zz}(\omega)$ are obtained
from eq. (62) by considering $\chi_{q}^{zz}(\omega-i\epsilon)$ in
the limit $\epsilon\longrightarrow0.$ We have the explicit result
\begin{equation}
Re\chi_{q}^{zz}\left(\omega\right)=-\frac{1}{N}P\sum_{k}\frac{\langle
n_{k}\rangle-\left\langle n_{k-q}\right\rangle
}{\omega+\epsilon_{k}-\epsilon_{k-q}},\end{equation} where $P$
denotes Cauchy principal value, and
\begin{equation}
Im\chi_{q}^{zz}\left(\omega\right)=\frac{\pi}{N}\sum\limits
_{k}(\left\langle n_{k}\right\rangle
-n_{k-q})\delta(\omega+\epsilon_{k}-\epsilon_{k-q}).\end{equation}
As in the isotropic model without long-range interactions on the
alternating superlattice, $\chi_{q}^{zz}\left(\omega\right)$ goes
to zero {[}23{]} for $\Gamma>\Gamma_{c}$ and for
$\Gamma<\Gamma_{c}$, and $Im\chi_{q}^{zz}\left(\omega\right)$ is
zero for $\omega>2J\sin(q/2).$ No significant differences in the
behaviour of the dynamic susceptibility are present in the
classical and quantum regions. There are no infinite singularities
in the imaginary part, and the finite discontinuities at the band
edges correspond to finite (lower edge) and infinite (upper edge)
divergences of the real part.
\section{{\large The classical-quantum crossover}}
The global phase diagram, in the space generated by temperature
$\widetilde{T}$ , field $\Gamma$ and ratio $r$ between short and
long-range interaction, is presented in Fig.9. The critical
surface has a mirror symmetry with respect to the plane
$\widetilde{T}\times r.$ As we can see, the classical second-order
critical line meets the quantum critical line at P$_{bc}.$
Therefore, P$_{bc}$ is a bicritical point as previously
conjectured {[}16{]} .
\begin{figure}[h]
\begin{center}
\includegraphics[width=16cm]{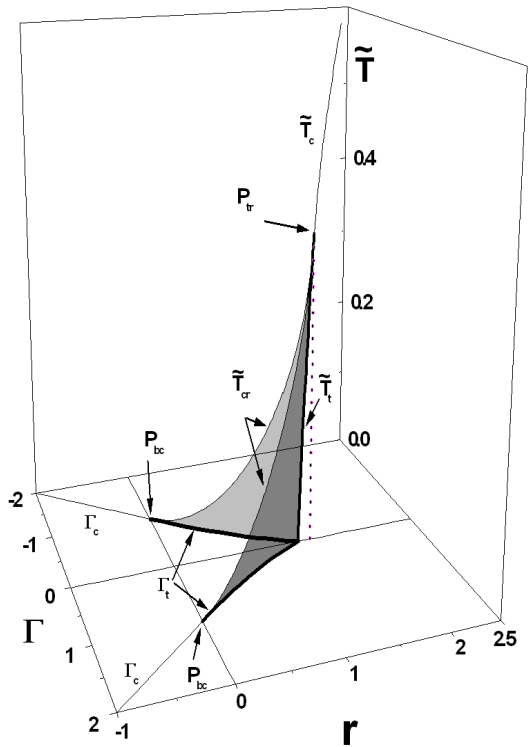}
\end{center}
\caption{Global phase diagram of the model as a function of
temperature $\widetilde{T}$ $(\widetilde{T}=k_{B}T/J)$, $\Gamma$
($\Gamma=h/J)$ and $r$ ($r=I/J$). $\widetilde{T}_{t}$ and
$\widetilde{T}_{c}$ identify the first and second-order classical
transitions, $\widetilde{T}_{cr}$ the classical critical line at
nonzero field, P$_{tr}$\ the classical tricritical point.
$\Gamma_{t}$ and\ $\Gamma_{c}$ identify the first and second-order
quantum transitions and P$_{bc}$the quantum bicritical point. }
\label{fig9}
\end{figure}
As shown in Fig. 5(b), there is a universal curve for the magnetization,
at $T=0$, in terms of the the effective field $\Gamma_{eff}$ ($\Gamma_{eff}\equiv\Gamma+2M^{z}r).$
This gives support to the conclusion that $\Gamma_{eff}$ is the relevant
variable, as far as the critical behaviour is concerned. Therefore,
the critical line can be drawn as a function of $\Gamma_{eff}.$ It
will end at the quantum critical point, $\Gamma_{eff}=1,$ for the
second-order transitions. Similarly, the classical first-order transition
line will end up at $\Gamma_{eff}=4/\pi,$ for $\Gamma=0,$ and the
critical surface can be collapsed into two (two-dimensional) diagrams
respectively related to the first and second-order transitions .

In these diagrams, which are characteristic of the quantum/classical
phase transitions, there are different regions where either classical
or quantum fluctuations dominate {[}24{]}. These regions are separated
by crossover lines which, in general, are not so easily defined, since
there is no unique criterion to characterize them {[}25{]}. In our
case, we have defined a criterion from the behaviour of the magnetization
as a function of temperature for different values of $\Gamma$, in
different critical regimes.

For $r$=-0.1, where we have a second-order quantum transition, the
magnetization as a function of the field for different values of $\Gamma$
is shown in Fig. 10(a). The main feature is the appearance of a peak
\ in the magnetization which disappears at the critical effective
field $\Gamma_{eff}=1,$ which corresponds in this case to $\Gamma=1.1.$
For values of the field larger than this value, the magnetization
is a decreasing monotonic function of temperature. In our view, this
means that the classical behaviour has been set in from zero temperature.
Therefore, we can associate the peaks in the magnetization with points
of the crossover line which, in this case, separates the critical
quantum and classical regimes.
\begin{figure}
\begin{center}
\includegraphics[width=14cm]{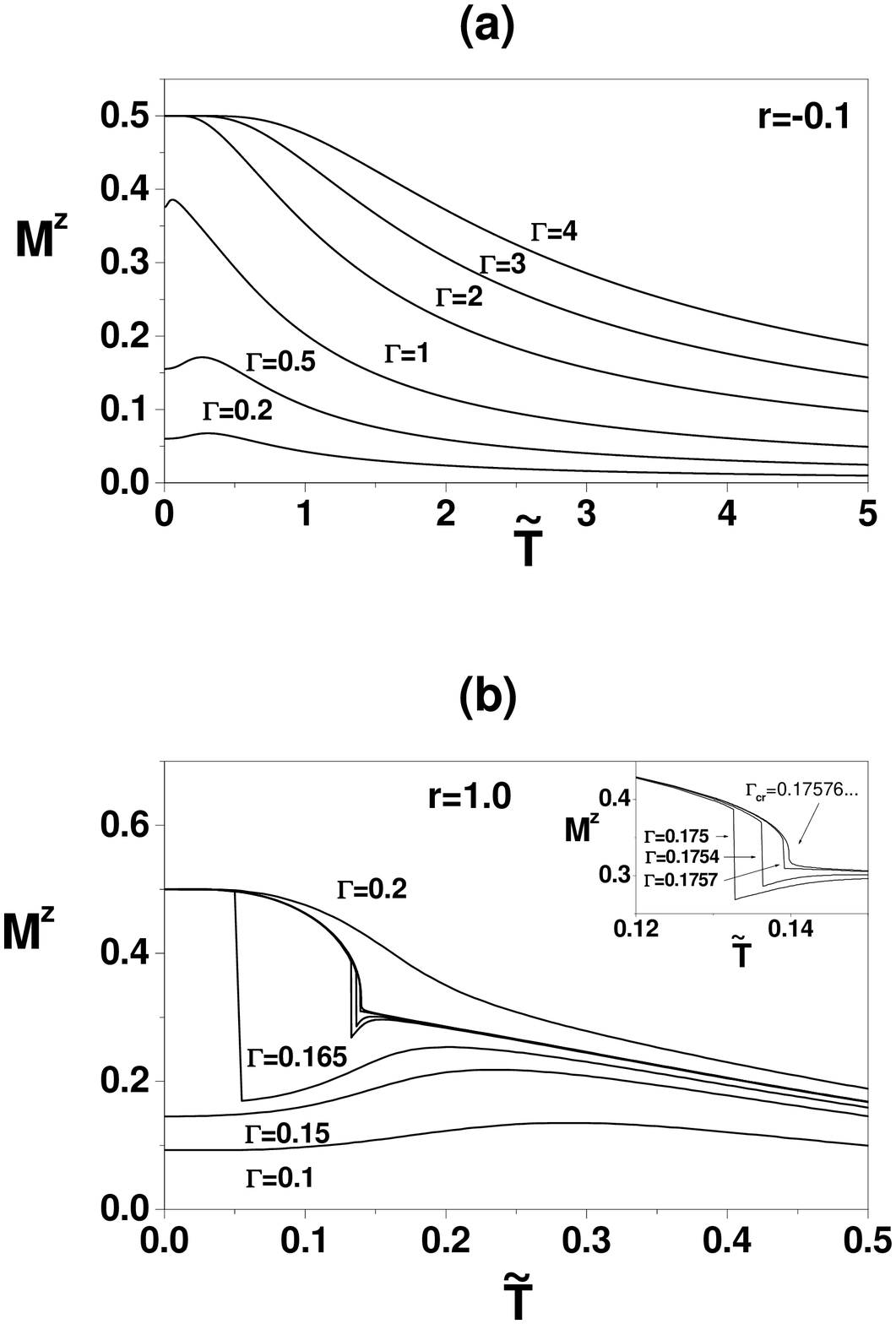}
\end{center}
\caption{Magnetization as a function of temperature
$\widetilde{T}\,\,(\widetilde{T}=k_{B}T/J)$ for different values
of the field. $\Gamma$($\Gamma=h/J)$. (a) For $r=-0.1$ $(r=I/J)$
and (b) for r$=1.0$.} \label{fig10}
\end{figure}
This critical classical/quantum diagram associated with the
second-order transitions is shown in Fig.11(a). Since, at
$\widetilde{T}=0,$ the system is in an ordered quantum state for
$\Gamma_{eff}<1,$ region I corresponds to the ordered quantum one
where the quantum fluctuations predominate. Region II corresponds
to the classical one where the classical fluctuations predominate.
For $\widetilde{T}=0$ and $\Gamma_{eff}>1,$ the system is in the
paramagnetic quantum state, but it becomes classical for
$\widetilde{T}\neq0.$ Consequently, this paramagnetic line
coincides with the crossover line.
For $r$=1.0, where we have a
first-order quantum transition, the magnetization as a function of
temperature presents a discontinuity in addition to the peak, as
shown in in Fig. 10(b). As it can be seen in the inset of the Fig.
10(b), the peak of the magnetization will disappear for a field
slightly smaller than the critical field \ $\Gamma_{c}.$\ This
means that the crossover does not coincide with the second-order
critical line shown in Fig. 11(a) as a function of $\Gamma_{eff},$
and ends at $\Gamma_{eff}=1.$ The first-order transition line ends
at $\Gamma_{eff}=4/\pi,$ which is the quantum transition point;
for $\Gamma_{eff}>4/\pi,$ the system is in the paramagnetic
quantum state. The points along the axis $\widetilde{T}=0,$\ for
$1<\Gamma_{eff}<4/\pi,$ are first-order quantum transitions
points; the classical/quantum critical diagram associated with the
first-order transitions is shown in Fig. 11(b). As in Fig.11(a),
region I corresponds to the ordered quantum state, where the
quantum fluctuations predominate, and region II to the classical
one, where the classical fluctuations predominate. For
$\Gamma_{eff}>1,$, as in the previous case, the crossover line
coincides with the axis $\widetilde{T}=0$ since the system becomes
classical for any $\widetilde{T}\neq0.$

\begin{figure}[h]
\begin{center}
\includegraphics[width=14cm]{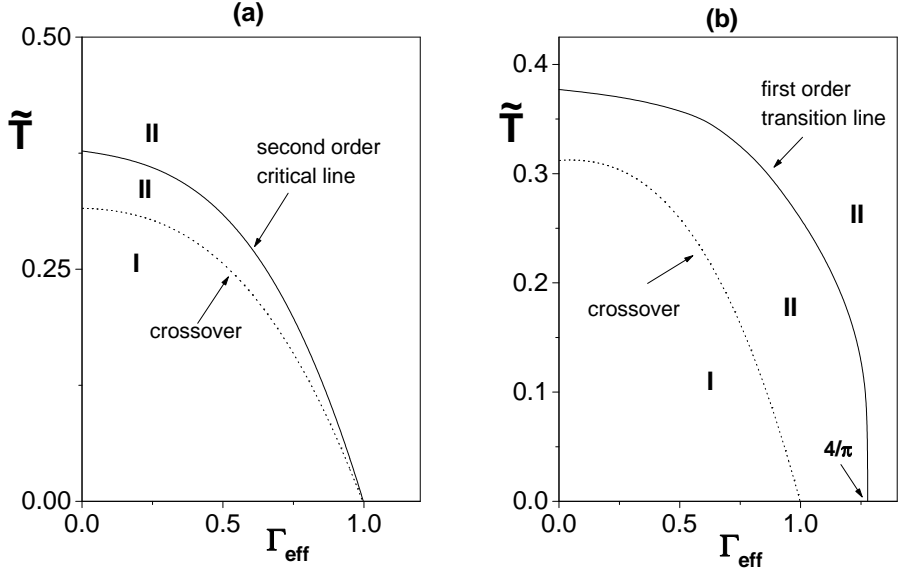}
\end{center}
\caption{Phase diagrams for first and second-order transitions as
function of $\Gamma_{eff}$ ($\Gamma=h/J,$ $r=I/J,$
$\Gamma_{eff}=\Gamma+2rM^{z})$. Region I corresponds to the
quantum regime and region II to the classical regime.
(a)Second-order transitions: for $\Gamma_{eff}>1,$ the crossover
line coincides with the paramagnetic quantum line. (b)First-order
transitions: for $\Gamma_{eff}>1,$ the crossover line coincides
with the first-order quantum points ($1<\Gamma_{eff}<4/\pi),$ and
with the paramagnetic quantum line ($\Gamma_{eff}>4/\pi).$}
\label{fig11}
\end{figure}

As we can see from Fig. 11, the phase diagrams for first and
second-order classical/quantum transition present a similar
structure. This new result gives support to the conjecture that
there is a universal structure of the phase diagram associated
with second-order quantum/classical transitions {[}1{]}.

\bigskip
\noindent{\large \textbf{Acknowledgements}}

The authors would like\ to thank the Brazilian agencies CNPq,
Capes and Finep for partial financial support. They would also
like to thank Dr. A. P. Vieira for useful discussions and a
critical reading of the manuscript.

\end{document}